\documentclass[twocolumn,prl,showpacs,superscriptaddress]{revtex4}%
\usepackage{amssymb}
\usepackage{amsmath}
\usepackage{amsfonts}
\usepackage{graphicx}%
\begin{document}
\title{All-Versus-Nothing Violation of Local Realism by Swapping Entanglement}
\author{Zeng-Bing Chen}
\affiliation{Physikalisches Institut, Universit\"{a}t Heidelberg, Philosophenweg 12,
D-69120 Heidelberg, Germany}
\affiliation{Hefei National Laboratory for Physical Sciences at Microscale and Department
of Modern Physics, University of Science and Technology of China, Hefei, Anhui
230026, China}
\author{Yu-Ao Chen}
\affiliation{Physikalisches Institut, Universit\"{a}t Heidelberg, Philosophenweg 12,
D-69120 Heidelberg, Germany}
\author{Jian-Wei Pan}
\affiliation{Physikalisches Institut, Universit\"{a}t Heidelberg, Philosophenweg 12,
D-69120 Heidelberg, Germany}
\affiliation{Hefei National Laboratory for Physical Sciences at Microscale and Department
of Modern Physics, University of Science and Technology of China, Hefei, Anhui
230026, China}

\begin{abstract}
The usual entanglement swapping protocol can entangle two particles that never
interact. Here we demonstrate an all-versus-nothing violation of local realism
for a partial entanglement swapping process. A Clauser-Horne-Shimony-Holt-type
inequality for two particles entangled by entanglement swapping is proposed
for a practical experiment and can be violated quantum mechanically up to $4$
beyond Cirel'son's bound $2\sqrt{2}$. Our result uncovers the intriguing
nonlocal character of entanglement swapping and may lead to a conclusive
(loophole-free) test of local realism versus quantum mechanics. The
experimental test of the nonlocality under current technology is also discussed.

\end{abstract}

\maketitle

By assuming local realism (LR), Einstein, Podolsky and Rosen (EPR) argued that
it seems possible to assign values with certainty to canonically conjugate
variables (e.g., position and momentum) if certain perfect quantum
correlations are used \cite{EPR}. EPR's observation is definitely in conflict
with quantum mechanics (QM), but remained for a long time untestable. In 1964,
Bell discovered his inequalities \cite{Bell,CHSH}, which since then enable
quantitative tests of QM versus LR for certain statistical correlations
predicted by QM. Strikingly, various versions of Bell's theorem without
inequalities\ have also been demonstrated
\cite{GHZ-89,GHZ-90,Mermin,Hardy,Pan-GHZ,Zhao-GHZ,Cabello,Chen-GHZ}.
Particularly, the contradiction between QM and local realistic theories arises
even for definite predictions, known as all-versus-nothing (AVN) violation
\cite{Mermin} of LR.

Entanglement is always an unavoidable ingredient in the violation of LR. In
various ways of creating entanglement, entanglement swapping
\cite{swap,swap-Pan} is perhaps most intriguing in the sense that it can
entangle two particles that never interact. In this paper, we demonstrate an
AVN violation of LR for a partial entanglement swapping process, where there
is no classical communication between different spatial locations and only
partial Bell-state measurement is required. A Clauser-Horne-Shimony-Holt
(CHSH) type inequality for two particles entangled by entanglement swapping is
proposed for a practical experiment and can be violated, up to $4$, by QM
beyond the standard Cirel'son bound \cite{bound} $2\sqrt{2}$ (One can also
surpass the bound by post-selected GHZ-entangled particles \cite{Cabello-c}).
The theorem proved here can lead to an ultimately loophole-free test of LR
against QM.

Without loss of generality and for definiteness, in the following we consider
polarized photons as example. The four Bell states of photons read $\left\vert
\psi^{\pm}\right\rangle =\frac{1}{\sqrt{2}}\left(  \left\vert H\right\rangle
\left\vert V\right\rangle \pm\left\vert V\right\rangle \left\vert
H\right\rangle \right)  $\ and $\left\vert \phi^{\pm}\right\rangle =\frac
{1}{\sqrt{2}}\left(  \left\vert H\right\rangle \left\vert H\right\rangle
\pm\left\vert V\right\rangle \left\vert V\right\rangle \right)  $, where
$\left\vert H\right\rangle $ ($\left\vert V\right\rangle $) stands for photons
with horizontal (vertical) polarization. They can be created via the
spontaneous parametric down-conversion in a nonlinear optical crystal
\cite{Kwiat} and satisfy (see, e.g., \cite{Cabello})
\begin{equation}%
\begin{array}
[c]{c}%
z\otimes z\left\vert \phi^{\pm}\right\rangle =\left\vert \phi^{\pm
}\right\rangle ,\ \ z\otimes z\left\vert \psi^{\pm}\right\rangle =-\left\vert
\psi^{\pm}\right\rangle ,\\
x\otimes x\left\vert \phi^{\pm}\right\rangle =\pm\left\vert \phi^{\pm
}\right\rangle ,\ \ x\otimes x\left\vert \psi^{\pm}\right\rangle
=\pm\left\vert \psi^{\pm}\right\rangle .
\end{array}
\label{zzxx}%
\end{equation}
Here $\sigma_{x}=\left\vert H\right\rangle \left\langle V\right\vert
+\left\vert V\right\rangle \left\langle H\right\vert \equiv x$ and $\sigma
_{z}=\left\vert H\right\rangle \left\langle H\right\vert -\left\vert
V\right\rangle \left\langle V\right\vert \equiv z$ are the Pauli-type
operators for polarizations of photons. Now one can also introduce another set
of Bell states $\left\vert \chi^{\pm}\right\rangle =\frac{1}{\sqrt{2}}\left(
\left\vert H\right\rangle \left\vert \bar{H}\right\rangle \pm\left\vert
V\right\rangle \left\vert \bar{V}\right\rangle \right)  $\ and\ $\left\vert
\omega^{\pm}\right\rangle =\frac{1}{\sqrt{2}}\left(  \left\vert V\right\rangle
\left\vert \bar{H}\right\rangle \pm\left\vert H\right\rangle \left\vert
\bar{V}\right\rangle \right)  $, with $\left\vert \bar{H}\right\rangle
=\frac{1}{\sqrt{2}}(\left\vert H\right\rangle +\left\vert V\right\rangle )$
and $\left\vert \bar{V}\right\rangle =\frac{1}{\sqrt{2}}(\left\vert
H\right\rangle -\left\vert V\right\rangle )$. Then one has
\begin{equation}%
\begin{array}
[c]{c}%
z\otimes x\left\vert \chi^{\pm}\right\rangle =\left\vert \chi^{\pm
}\right\rangle ,\ \ z\otimes x\left\vert \omega^{\pm}\right\rangle
=-\left\vert \omega^{\pm}\right\rangle ,\\
x\otimes z\left\vert \chi^{\pm}\right\rangle =\pm\left\vert \chi^{\pm
}\right\rangle ,\ \ x\otimes z\left\vert \omega^{\pm}\right\rangle
=\pm\left\vert \omega^{\pm}\right\rangle .
\end{array}
\label{zxzx}%
\end{equation}

Now let us consider the standard entanglement swapping protocol for
EPR-entangled photon pairs: EPR source I (II) \textit{independently} emits
maximally entangled photons 1 and 2 in state $\left\vert \psi^{-}\right\rangle
_{12}$ (3 and 4 in state $\left\vert \psi^{-}\right\rangle _{34}$); if photons
2 and 3 are subject to a Bell-state measurement, photons 1 and 4 will then be
maximally entangled conditioned on the outcomes of the Bell-state measurement.
This can easily be seen by writting down the initial state $\left\vert
\psi^{-}\right\rangle _{12}\left\vert \psi^{-}\right\rangle _{34}$ in terms of
the Bell states in either \{$\left\vert \psi^{\pm}\right\rangle ,\left\vert
\phi^{\pm}\right\rangle $\} or \{$\left\vert \chi^{\pm}\right\rangle
,\left\vert \omega^{\pm}\right\rangle $\}:
\begin{align}
\left\vert \psi^{-}\right\rangle _{12}\left\vert \psi^{-}\right\rangle _{34}
&  =\frac{1}{2}(\left\vert \psi^{-}\right\rangle _{14}\left\vert \psi
^{-}\right\rangle _{23}-\left\vert \psi^{+}\right\rangle _{14}\left\vert
\psi^{+}\right\rangle _{23}\nonumber\\
&  +\left\vert \phi^{+}\right\rangle _{14}\left\vert \phi^{+}\right\rangle
_{23}-\left\vert \phi^{-}\right\rangle _{14}\left\vert \phi^{-}\right\rangle
_{23})\nonumber\\
&  =\frac{1}{2}(\left\vert \omega^{+}\right\rangle _{14}\left\vert \omega
^{+}\right\rangle _{23}-\left\vert \omega^{-}\right\rangle _{14}\left\vert
\omega^{-}\right\rangle _{23}\nonumber\\
&  -\left\vert \chi^{+}\right\rangle _{14}\left\vert \chi^{+}\right\rangle
_{23}+\left\vert \chi^{-}\right\rangle _{14}\left\vert \chi^{-}\right\rangle
_{23}). \label{bsm}%
\end{align}
Then using the properties of the Bell states in (\ref{zzxx}) and (\ref{zxzx})
one has the following eigenequations
\begin{align}
z_{1}\cdot z_{2}z_{3}\cdot z_{4}\left\vert \psi^{-}\right\rangle
_{12}\left\vert \psi^{-}\right\rangle _{34}  &  =\left\vert \psi
^{-}\right\rangle _{12}\left\vert \psi^{-}\right\rangle _{34},\label{e1}\\
x_{1}\cdot x_{2}x_{3}\cdot x_{4}\left\vert \psi^{-}\right\rangle
_{12}\left\vert \psi^{-}\right\rangle _{34}  &  =\left\vert \psi
^{-}\right\rangle _{12}\left\vert \psi^{-}\right\rangle _{34},\label{e2}\\
z_{1}\cdot z_{2}x_{3}\cdot x_{4}\left\vert \psi^{-}\right\rangle
_{12}\left\vert \psi^{-}\right\rangle _{34}  &  =\left\vert \psi
^{-}\right\rangle _{12}\left\vert \psi^{-}\right\rangle _{34},\label{e3}\\
x_{1}\cdot x_{2}z_{3}\cdot z_{4}\left\vert \psi^{-}\right\rangle
_{12}\left\vert \psi^{-}\right\rangle _{34}  &  =\left\vert \psi
^{-}\right\rangle _{12}\left\vert \psi^{-}\right\rangle _{34}. \label{e4}%
\end{align}

Now suppose that the measurements on photon 1, photons 2 and 3, and photon 4
are performed on three regions that are mutually spacelike separated. Then the
perfect correlations in Eqs. (\ref{e1})-(\ref{e4}) allow one to infer that
operators separated by $(\cdot)$ can be identified as EPR's local
\textquotedblleft elements of reality\textquotedblright\ (LERs). For instance,
by measuring $z_{1}$ and $z_{4}$ one can predict with certainty the value of
$z_{2}z_{3}$ without in any way disturbing photons 2 and 3, and as such
$z_{2}z_{3}$\ can be regarded as an LER according to the very notion of the
EPR local realism. Similar reasoning shows that one may establish the
following LERs: $(z_{1},x_{1})$ for photon 1, $\left(  z_{2}z_{3},x_{2}%
x_{3},z_{2}x_{3},x_{2}z_{3}\right)  $ for photons 2 and 3, and $\left(
z_{4},x_{4}\right)  $ for photon 4. A local realistic interpretation of the
quantum results (\ref{e1})-(\ref{e4}) is to assume that the individual value
of any local operator $\left(  z_{1},x_{1};z_{2}z_{3},x_{2}x_{3},z_{2}%
x_{3},x_{2}z_{3};z_{4},x_{4}\right)  $ is predetermined. These predetermined
values are denoted by $v\left(  V\right)  =\pm1$ for the observable
$V\in\left(  z_{1},x_{1};z_{2}z_{3},x_{2}x_{3},z_{2}x_{3},x_{2}z_{3}%
;z_{4},x_{4}\right)  $. To be consistent with Eqs. (\ref{e1})-(\ref{e4}), LR
then predicts
\begin{align}
v\left(  z_{1}\right)  v\left(  z_{2}z_{3}\right)  v\left(  z_{4}\right)   &
=+1,\ \ v\left(  x_{1}\right)  v\left(  x_{2}x_{3}\right)  v\left(
x_{4}\right)  =+1,\nonumber\\
v\left(  z_{1}\right)  v\left(  z_{2}x_{3}\right)  v\left(  x_{4}\right)   &
=+1,\ \ v\left(  x_{1}\right)  v\left(  x_{2}z_{3}\right)  v\left(
z_{4}\right)  =+1.\label{lhv}%
\end{align}
Unfortunately, the relations in Eq. (\ref{lhv}) do not show any conflict
between QM and LR.

The missed link for demonstrating the conflict can be uncovered by noting that
one also has a \textit{quantum identity}
\begin{equation}
z_{2}z_{3}\cdot x_{2}x_{3}=-z_{2}x_{3}\cdot x_{2}z_{3}, \label{id}%
\end{equation}
regardless of the states of photons 2 and 3. To reproduce the above identity
local realistic theories must also predict that
\begin{equation}
v\left(  z_{2}z_{3}\right)  v\left(  x_{2}x_{3}\right)  =\epsilon
=\pm1,\ v\left(  z_{2}x_{3}\right)  v\left(  x_{2}z_{3}\right)  =-\epsilon.
\label{lhvid}%
\end{equation}

Now it is ready to see that Eqs. (\ref{lhv}) and (\ref{lhvid}) are mutually
inconsistent: Multiplying the left-hand sides of the relations in Eqs.
(\ref{lhv}) and (\ref{lhvid}) gives $+1$ (Note that $v\left(  V\right)  =\pm
1$), whereas the product of the right-hand sides yields $-1$. Thus we do have
a GHZ-type contradiction for the perfect correlations in the entanglement
swapping process. Namely, LR predicts that the above LERs must satisfy five of
the six relations in Eqs. (\ref{lhv}) and (\ref{lhvid}) so that the
corresponding five quantum predictions can be reproduced. Then on the level of
\textit{gedanken} experiment, every measurement of the LERs related with the
sixth relations must give a perfect correlation, which is, however, exactly
opposite to the sixth quantum prediction. In other words, certain outcomes
predicted to definitely occur by LR are never allowed to occur by QM and vice
versa. This completes the demonstration of an AVN violation of LR for the
perfect correlations in an ideal entanglement swapping.

A careful reader might have found that the same operators may appear in
different equations in (\ref{e1})-(\ref{e4}) and (\ref{id}). For instance,
$z_{2}z_{3}$ and $x_{2}x_{3}$ not only appear separately in Eqs. (\ref{e1})
and (\ref{e2}), but also appear jointly in Eq. (\ref{id}) [see also
(\ref{plus}) and (\ref{minus})]. In order to validate our AVN nonlocality
argument, it is, however, necessary to assign always a single value to the
same operator, though it can appear in different equations. Therefore, one
either has to assume noncontexuality (e.g., measurement of $z_{2}z_{3}$ does
not disturb the value of $x_{2}x_{3}$ and vice versa) or one has to be able to
measure $z_{2}z_{3}$, $x_{2}x_{3}$ and $z_{2}z_{3}\cdot x_{2}x_{3}$ with the
same apparatuses so that they appear only in the same experimental context.

Therefore, to establish the observables in two groups $(z_{2}z_{3},x_{2}%
x_{3})$ and $(z_{2}x_{3},x_{2}z_{3})$\ as LERs, the two observables of each
group must be measured by one and the same apparatus (Note that this is
possible because the two observables of each group commute with each other).
Explicitly, our argument requires two complementary apparatuses $A$ and $B$:
Apparatus A (B) measures $z_{2}z_{3}$\ and $x_{2}z_{3}$\ ($z_{2}x_{3}$ and
$x_{2}z_{3}$) simutaneously, and by multiplication gets the value of
$z_{2}z_{3}\cdot x_{2}x_{3}$\ ($z_{2}x_{3}\cdot x_{2}z_{3}$). This eliminates
the necessity of a noncontexuality assumption, similarly to an AVN nonlocality
argument for doubly-entangled two particles \cite{Chen-GHZ}.

However, there is still another problem that might invalidate our argument:
Though Eq. (\ref{id}) is a quantum identity, there is no \textit{a priori}
reason why each individual measurement result of $z_{2}z_{3}\cdot x_{2}x_{3}$
must be opposite to that of $z_{2}x_{3}\cdot x_{2}z_{3}$ (Of course, the mean
value of $z_{2}z_{3}\cdot x_{2}x_{3}$ must be exactly opposite to that of
$z_{2}x_{3}\cdot x_{2}z_{3}$). This is because the collapse to any eigenstate
of $z_{2}z_{3}\cdot x_{2}x_{3}$\ (or $z_{2}x_{3}\cdot x_{2}z_{3}$) is
completely random according to QM. Thus, for the above argument to hold, one
must ensure that the individual measurement results of $z_{2}z_{3}\cdot
x_{2}x_{3}$ and $z_{2}x_{3}\cdot x_{2}z_{3}$\ are exactly opposite.

Surprisingly enough, one can indeed achieve this by introducing
\begin{align}
\left\vert \epsilon=\pm1\right\rangle _{1234}  &  \equiv\frac{1}{\sqrt{2}%
}\left(  \left\vert \psi^{\mp}\right\rangle _{14}\left\vert \psi^{\mp
}\right\rangle _{23}+\left\vert \phi^{\pm}\right\rangle _{14}\left\vert
\phi^{\pm}\right\rangle _{23}\right) \nonumber\\
&  \equiv\frac{1}{\sqrt{2}}\left(  \left\vert \chi^{\mp}\right\rangle
_{14}\left\vert \chi^{\mp}\right\rangle _{23}+\left\vert \omega^{\pm
}\right\rangle _{14}\left\vert \omega^{\pm}\right\rangle _{23}\right)  ,
\label{new}%
\end{align}
with $\left\vert \psi^{-}\right\rangle _{12}\left\vert \psi^{-}\right\rangle
_{34}=\frac{1}{\sqrt{2}}\left(  \left\vert \epsilon=+1\right\rangle
_{1234}-\left\vert \epsilon=-1\right\rangle _{1234}\right)  $. Importantly, in
terms of $\left\vert \epsilon=\pm1\right\rangle _{1234}$, Eqs. (\ref{e1}%
)-(\ref{e4}) are still valid, but with $\left\vert \psi^{-}\right\rangle
_{12}\left\vert \psi^{-}\right\rangle _{34}$ being replaced by either
$\left\vert \epsilon=+1\right\rangle _{1234}$\ or $\left\vert \epsilon
=-1\right\rangle _{1234}$. Moreover, one also has
\begin{align}
z_{2}z_{3}\cdot x_{2}x_{3}\left\vert \epsilon=\pm1\right\rangle _{1234}  &
=\pm\left\vert \epsilon=\pm1\right\rangle _{1234},\label{plus}\\
z_{2}x_{3}\cdot x_{2}z_{3}\left\vert \epsilon=\pm1\right\rangle _{1234}  &
=\mp\left\vert \epsilon=\pm1\right\rangle _{1234}. \label{minus}%
\end{align}
Therefore, if one can select either $\left\vert \epsilon=+1\right\rangle
_{1234}$\ or $\left\vert \epsilon=-1\right\rangle _{1234}$\ out of the initial
state $\left\vert \psi^{-}\right\rangle _{12}\left\vert \psi^{-}\right\rangle
_{34}$, then the above AVN nonlocality argument remains to be applicable to
the selected $\left\vert \epsilon=\pm1\right\rangle _{1234}$. This selection
is equivalent to the preparation of the state $\left\vert \epsilon
=+1\right\rangle _{1234}$\ or $\left\vert \epsilon=-1\right\rangle _{1234}$.

As a mathematical proof and at the level of \textit{gedanken} experiment, the
above AVN proof works as it is. However, a real experiment always faces
imperfect correlations and nonideal detections. To overcome this difficulty, a
Bell-type inequality for statistical correlations in the present context is required.

To this end, one should note the following facts:
\begin{align}
v\left(  z_{2}z_{3}\right)   &  \equiv m=\pm1\Leftrightarrow v\left(
x_{2}x_{3}\right)  =\epsilon m,\nonumber\\
v\left(  z_{2}x_{3}\right)   &  \equiv n=\pm1\Leftrightarrow v\left(
x_{2}z_{3}\right)  =-\epsilon n, \label{mn}%
\end{align}
which are the consequences of Eq. (\ref{lhvid}). Then from Eqs. (\ref{e1}%
)-(\ref{e4}), where $\left\vert \psi^{-}\right\rangle _{12}\left\vert \psi
^{-}\right\rangle _{34}$ is now replaced by either $\left\vert \epsilon
=+1\right\rangle _{1234}$\ or $\left\vert \epsilon=-1\right\rangle _{1234}$,
one can prove the following CHSH-type inequality, which must be fulfilled by
any local realistic theory, for the two entanglement-swapped photons 1 and 4:
\begin{equation}
\left\vert \left\langle mz_{1}z_{4}+nz_{1}x_{4}-\epsilon nx_{1}z_{4}+\epsilon
mx_{1}x_{4}\right\rangle _{LR}\right\vert \leq2. \label{chsh}%
\end{equation}
Actually,
\begin{align}
\mathcal{LR}  &  \equiv\left\vert \left\langle mz_{1}z_{4}+nz_{1}%
x_{4}-\epsilon nx_{1}z_{4}+\epsilon mx_{1}x_{4}\right\rangle _{LR}\right\vert
\nonumber\\
&  =\left\vert \left\langle z_{1}z_{4}+mnz_{1}x_{4}-\epsilon mnx_{1}%
z_{4}+\epsilon x_{1}x_{4}\right\rangle _{LR}\right\vert \nonumber\\
&  =\left\vert \left\langle z_{1}\left(  z_{4}+mnx_{4}\right)  -\epsilon
x_{1}\left(  mnz_{4}-x_{4}\right)  \right\rangle _{LR}\right\vert \nonumber\\
&  =\left\{
\begin{array}
[c]{c}%
\left.  \left\vert \left\langle z_{1}\left(  z_{4}+x_{4}\right)  -\epsilon
x_{1}\left(  z_{4}-x_{4}\right)  \right\rangle _{LR}\right\vert
,\ (mn=1)\right. \\
\left\vert \left\langle z_{1}\left(  z_{4}-x_{4}\right)  +\epsilon
x_{1}\left(  z_{4}+x_{4}\right)  \right\rangle _{LR}\right\vert ,\ (mn=-1)
\end{array}
\right.  \label{chsh4}%
\end{align}
which are the Bell correlation functions in the present context and satisfy
$\mathcal{LR}\leq2$.

It should be noted that Eq. (\ref{chsh}) is, in fact, the local realistic
prediction for $\mathcal{M}=z_{1}\cdot z_{2}z_{3}\cdot z_{4}+x_{1}\cdot
x_{2}x_{3}\cdot x_{4}+z_{1}\cdot z_{2}x_{3}\cdot x_{4}+x_{1}\cdot x_{2}%
z_{3}\cdot z_{4}$, namely,
\begin{equation}
\left\vert \left\langle \mathcal{M}\right\rangle _{LR}\right\vert
\leq2\label{mlr}%
\end{equation}
under the restriction of Eq. (\ref{id}) or (\ref{lhvid}). However, quantum
prediction of $\mathcal{M}$ with respect to $\left\vert \epsilon
=+1\right\rangle _{1234}$\ or $\left\vert \epsilon=-1\right\rangle _{1234}%
$\ can be as large as $4$, i.e.,
\begin{equation}
\left\vert \left\langle \mathcal{M}\right\rangle _{QM}\right\vert
\leq4,\label{mqm}%
\end{equation}
which is beyond the Cirel'son bound $2\sqrt{2}$. The upper bound can be
reached by either $\left\vert \epsilon=+1\right\rangle _{1234}$\ or
$\left\vert \epsilon=-1\right\rangle _{1234}$. This \textit{much larger
violation of the CHSH inequality }(\ref{mlr})\textit{ by the quantum
prediction }(\ref{mqm})\textit{ shows more resistance of }(\ref{mlr})\textit{
to noise and the maximal visibility to fulfil }(\ref{mlr})\textit{ can be
reduced to }$50\%$\textit{.} In this kind of the Bell experiment testing
(\ref{mlr}) the detection efficiency for photons 2 and 3 does not contribute
to the overall detection efficiency because the proposed experiment can be
regarded as an \textquotedblleft event-ready\textquotedblright\ Bell test
\cite{swap}.

The above feature of our CHSH inequality (\ref{mlr}) is very helpful to close
the detection-efficiency loophole in the Bell-type experiments (The
detection-efficiency loophole has been closed in \cite{detection} by using
entangled ions, which can be detected with nearly perfect efficiency). As
entangled photons are ideal for closing the locality loophole, as realized in
Ref. \cite{Weihs}, our result opens the door for a loophole-free test of LR.
In this context, we can use two independent ion-photon pairs (all in the state
$\left\vert \psi^{-}\right\rangle $; see, e.g.,
\cite{ion-photon-TH,ion-photon-EX}), where the two ions are located at two
remote locations and the two photons in the third remote location are subject
to measurements as in Eqs. (\ref{e1})-(\ref{e4}), (\ref{plus}) and
(\ref{minus}). A recent experiment reported the required ion-photon
entanglement with fidelity $\approx$0.87 \cite{ion-photon-EX}, showing that
the loophole-free event-ready\ Bell test is feasible under current technology.

The inequality (\ref{mlr}) implies that, if entanglement swapping could be
interpreted with LR by identifying $(z_{2}z_{3},x_{2}x_{3})$ and $(z_{2}%
x_{3},x_{2}z_{3})$\ as EPR's elements of reality, then LR predicts further
four CHSH inequalities in (\ref{chsh4}). However, these inequalities can be
quantum mechanically violated up to 4, though photons 1 and 4 do not interact
in any way and even have no common history. Thus, the entanglement swapping
channel can violate LR in an intriguing way. As noted by Aspect,
\textquotedblleft This would certainly help us to further understand
nonlocality\textquotedblright\ \cite{Aspect}.%
\begin{figure}
[ptb]
\begin{center}
\includegraphics[
height=1.9865in,
width=2.4319in
]%
{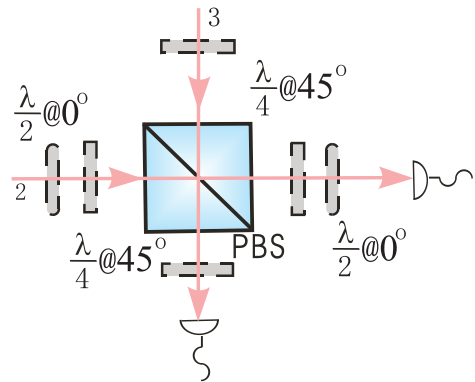}%
\caption{A linear optics apparatus selecting $\left\vert \epsilon
=+1\right\rangle _{1234}$\ out of $\left\vert \psi^{-}\right\rangle
_{12}\left\vert \psi^{-}\right\rangle _{34}$. The four quarter wave plates (at
45$^{\circ}$) can affect the transformations $R=\frac{1}{\sqrt{2}%
}(H+iV)\rightarrow H$ and $L=\frac{1}{\sqrt{2}}(H-iV)\rightarrow V$; the two
half wave plates function as $R\leftrightarrow L$; the polarizing beam
splitter (PBS) reflects $V$ photons and transmits $H$ photons. Only for
$\left\vert \epsilon=+1\right\rangle _{1234}$ there will be a coincidence
detection between the two detectors.}%
\end{center}
\end{figure}

In our GHZ-type theorem for swapping entanglement, we made use of two
independent entanglement sources. By taking into account the independence of
the two sources, we can further prove \cite{Chen} that the counts for the
experiments in (\ref{e3}) and (\ref{e4}) must be null according to LR, for any
nonzero detection efficiency. Quantum mechanically, the counts for the
experiments in (\ref{e3}) and (\ref{e4}) depend on the efficiency of the used
detectors and are of course nonzero. Thus, the proved theorem can lead to a
loophole-free test of LR against QM even using detectors with arbitrarily low
efficiency. It is worthwhile to point out that our AVN argument is valid
regardless of whether the two entanglement sources are independent or not. The
two apparatuses A and B used to avoid the contextuality assumption are crucial
in this aspect.

Finally, let us discuss briefly the practical aspects of implementing the
required measurements. Measuring jointly the two observables in $(z_{2}%
z_{3},x_{2}x_{3})$ [$(z_{2}x_{3},x_{2}z_{3})$] is equivalent \cite{Cabello} to
making a complete discrimination between four Bell states \{$\left\vert
\psi^{\pm}\right\rangle _{23},\left\vert \phi^{\pm}\right\rangle _{23}$\}
(\{$\left\vert \chi^{\pm}\right\rangle _{23},\left\vert \omega^{\pm
}\right\rangle _{23}$\}). In the case of polarized photons, such a complete
Bell-state measurement is impossible with only linear optics and necessitates
nonlinear optical interactions at the single-photon level \cite{impossible}.
Fortunately, a nondestructive controlled-NOT ({\small CNOT}) gate for two
independent photons using only linear optical elements is sufficient for the
present purpose and has been realized in a very recent experiment
\cite{Zhao-CNOT}. This experimentally demonstrated {\small CNOT} gate is
probabilistic, but \textquotedblleft event-ready\textquotedblright\ in the
sense that one can detect when the gate has succeeded by performing some
appropriate measurement on ancilla photons. The information gathered in the
measurement can then be feed-forwarded for conditional future operations on
the photonic qubits. As pointed out by Cabello \cite{Cabello}, distinguishing
between the two results $z_{2}z_{3}\cdot x_{2}x_{3}=\pm1$\ (or $z_{2}%
x_{3}\cdot x_{2}z_{3}=\pm1$) corresponds to partial Bell-state measurement,
which can be accomplished by only linear optical elements
\cite{impossible,BSM-lo}. In Fig. 1, we propose a linear optics apparatus
which can select $\left\vert \epsilon=+1\right\rangle _{1234}$\ out of
$\left\vert \psi^{-}\right\rangle _{12}\left\vert \psi^{-}\right\rangle _{34}%
$: There will be one photon in each output mode only for $\left\vert
\epsilon=+1\right\rangle _{1234}$, which can then be subject to future
measurements. For $\left\vert \epsilon=-1\right\rangle _{1234}$ photons 2 and
3 will go together to one of the two output modes. In this way, coincidence
detection between the two output modes, after all necessary measurements being
performed on photons 2 and 3, may distinguish $\left\vert \epsilon
=+1\right\rangle _{1234}$\ (corresponding to a coincidence detection) from
$\left\vert \epsilon=-1\right\rangle _{1234}$ (corresponding to no coincidence detection).

In summary, we have demonstrated an AVN violation of LR for a partial
entanglement swapping process. A CHSH inequality for two particles entangled
by entanglement swapping is proposed for a practical experiment and can be
violated, up to $4$, by QM. This larger violation is helpful to close the
detection-efficiency loophole in the Bell-type experiments for photons. For
two turely independent entanglement sources, our argument may lead to a
conflict between LR and QM for detectors with any nonzero efficiency. Compared
with the usual nonlocality argument, our result uncovers the intriguing
nonlocality of entanglement swapping and provides a theoretical base for an
ultimately loophole-free refutation of local realism. The experimental test of
the nonlocality under current technology is also discussed by using linear
optical elements.

This work was supported by the National NSF of China, the Fok Ying Tung
Education Foundation, the Chinese Academy of Sciences and the National
Fundamental Research Program. We also acknowledge the support by the European
Commission under Contract No. 509487 (Marie Curie Fellowship).

\textit{Note added}--After the completion of our work, we became aware of the
beautiful papers by Greenberger, Horne, and Zeilinger \cite{GHZ-new} who
obtained a similar result, without touching the contextuality issue. Hail the
GHZ factorization theorem!

\end{document}